# SIMBOL–X : a new generation hard X-ray telescope

P. Ferrando[*a], M. Arnaud[a], B. Cordier[a], A. Goldwurm[a], O. Limousin[a], J. Paul[a], J.L. Sauvageot[a],
P.O. Petrucci[b], M. Mouchet[c], G. Bignami[d], O. Citterio[e], S. Campana[e], G. Pareschi[e],
G. Tagliaferri[e], U. Briel[f], G. Hasinger[f], L. Strueder[f], P. Lechner[g], E. Kendziorra[h], M. Turner[i]

[a] DSM/DAPNIA/Service d'Astrophysique, CEA/Saclay, 91191 Gif-sur-Yvette Cedex, France
[b] Laboratoire d'Astrophysique de l'Observatoire de Grenoble, 38041 Grenoble Cedex, France
[c] LUTH, Observatoire de Paris Meudon, 92195 Meudon Cedex, France
[d] Centre d'Etude Spatiale des Rayonnements, BP 4346, 31028 Toulouse Cedex, France
[e] Observatorio Astronomico di Brera, 23807 Merate (Lc), Italy
[f] Max Planck Institut für Extraterrestrische Physik, 85748 Garching, Germany
[g] PNSensor GmbH, Roemerstr. 28, D-80803 München, Germany
[h] Institut für Astronomie and Astrophysik, 72076 Tübingen, Germany
[i] Dept of Astronomy and Astrophysics, Leicester Univ., Leicester LE1 7RH, United Kingdom

**ABSTRACT**

SIMBOL-X is a hard X–ray mission, operating in the 0.5–70 keV range, which is proposed by a consortium of European laboratories for a launch around 2010. Relying on two spacecraft in a formation flying configuration, SIMBOL–X uses a 30 m focal length X–ray mirror to achieve an unprecedented angular resolution (30 arcsec HEW) and sensitivity (100 times better than INTEGRAL below 50 keV) in the hard X–ray range. SIMBOL–X will allow to elucidate fundamental questions in high energy astrophysics, such as the physics of accretion onto Black Holes, of acceleration in quasar jets and in supernovae remnants, or the nature of the hard X–ray diffuse emission. The scientific objectives and the baseline concepts of the mission and hardware design are presented.

## 1. INTRODUCTION

The study of the non-thermal component in high energy astrophysics sources is presently hampered by the large gap in spatial resolution and sensitivity between the X–ray and gamma–ray domains. Below ~ 10 keV, astrophysics missions like XMM–Newton and Chandra are using X–ray mirrors based on grazing incidence reflection properties. This allows to achieve an extremely good angular resolution, down to 0.5 arcsec for Chandra, and a good signal to noise thanks to the focusing of the X–rays onto a small detector surface. This technique has however an energy limitation at ~ 10 keV due in particular to the maximum focal length that can fit in a single spacecraft. Hard X–ray and gamma–ray imaging instruments, such as those on the recently launched INTEGRAL are thus using a different technique, that of coded masks. This non-focusing technique has intrinsically a much lower signal to noise ratio than that of a focusing instrument, and does not allow to reach angular resolutions better than a few arc minutes. In addition to the difference in angular resolution, there is also roughly two orders of magnitude of difference in point source sensitivity between X–ray and gamma–ray telescopes.

This transition of techniques unfortunately happens roughly at the energy above which the identification of a non-thermal component is unambiguous with respect to thermal emission. Considered from the low energy side, this obviously strongly limits the interpretation of the high quality X–ray measurements, and particularly that related to the acceleration of particles. Considered from the high energy side, this renders impossible the mapping of the gamma–ray emission of extended sources to the scales needed to understand the emission mechanisms by comparing with lower energy data, and this limits the studies to very bright sources only.

---

[*] Further author information: send correspondence to Philippe Ferrando, e-mail: pferrando@cea.fr

The proposed SIMBOL–X will bridge this gap by extending the X–ray focusing technique to much higher energies, up to ~ 70 keV in its nominal design. In addition to that, with an energy range starting at ~ 0.5 keV SIMBOL–X will fully cover the transition from thermal to non-thermal emissions, as well as the Iron line region, two important characteristics for the study of the highly variable accreting sources which are prime scientific targets of this mission.

## 2. SCIENTIFIC OBJECTIVES

Offering "soft X–ray„-like angular resolution and sensitivity over its full energy range, SIMBOL–X will provide unique clues to investigate the physics of accretion around Black Holes, to elucidate the origin of the diffuse X–ray background above 10 keV and of the hard emission of the Galactic Centre, as well as to study the limits of particle acceleration in extended sources such as SNRs. SIMBOL–X will also provide clues to other high energy problems, as e.g. the origin of the non-thermal emission in clusters of galaxies. We shortly review here some of these objectives in correspondence with SIMBOL–X capabilities.

**2.1 Accretion onto compact objects**
By providing the most extreme gravitational conditions under which matter can be presently observed, the Black Hole environment is a unique laboratory to test laws of physics. If it is generally admitted that the accretion of matter onto these objects occurs via a very dynamic accretion disk, the way that this disk evolves, the mechanisms that form the often observed relativistic jets, as well as the origin of the high energy radiation remain unclear. SIMBOL–X will allow to perform unique investigations both for the supermassive Black Holes residing in the centre of Active Galactic Nuclei, and for the stellar mass Black Holes in our Galaxy and in our local group.

An example of the power of SIMBOL–X is illustrated in Figure 1 that shows a simulation of a 5 ks observation of an AGN, NGC 5548, which showed a wide Iron line in 1996 and a narrow one in 2000[1]. In this very short observation time, SIMBOL–X will give a very accurate measurement of the Iron line, the primary continuum, and the reflection bump. The full information, obtained simultaneously, is required to determine the precise shape of the Iron line, as well as to compare its variability with that of the continuum. This unique capability will be crucial to study the puzzling very rapid variability (less than a day) of the narrow Iron line which has been observed in several cases[2,3,4,5].

On X–ray Binaries, both of Low and High Mass, SIMBOL–X will allow to measure very detailed spectra in a short integration time, and to follow their evolution with time. This will give unique clues to the geometry of these systems, and on the problem of the origin of the high energy emission, for which models involving synchrotron emission from compact jets challenge the more common model of inverse Compton processes in a hot corona. Thanks to the very low background of SIMBOL–X, it will also be possible to observe these systems when they are in quiescence, i.e. a completely unknown regime of very weak accretion rate, and to perform very high signal to noise studies on QPO's at high energy, shedding light on the geometry of the flow very close to the central engine.

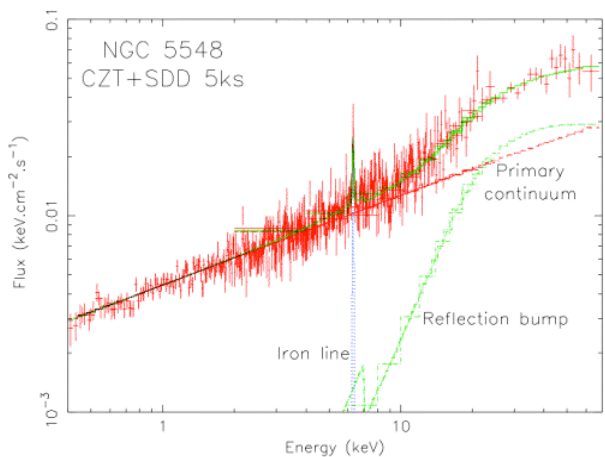
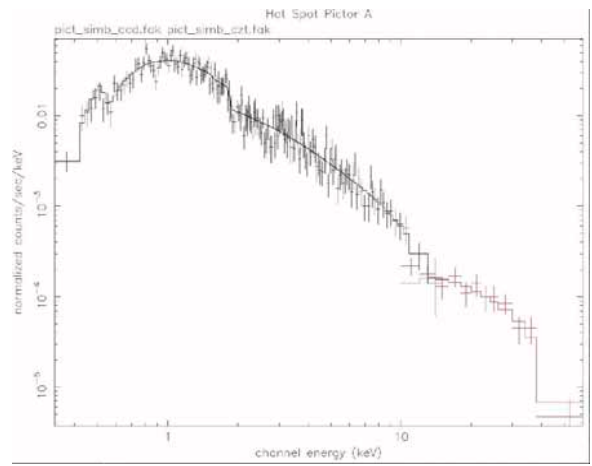

Figure 1 : Simulation of a 5 ks observation of NGC 5548 spectrum with SIMBOL–X.

Figure 2 : Simulation of a 50 ks observation of the spectrum of Pictor A hot spot.

Finally, the angular resolution of SIMBOL–X will allow to map the high energy emission of extended X–ray jets, like those of Cen A and Pictor A, which above 10 keV cannot be separated today from the bright emission of the accreting object. This is crucial to understand the emission mechanisms at work in these jets (synchrotron, inverse Compton), and in case of synchrotron emission to measure the maximum energy of the accelerated electrons. Figure 2 shows for example a simulated observation of the Pictor A jet hot spot, located 4 arcmins from the active nucleus, assuming that the power law shape observed by Chandra[6] extends to very high energy. With this 50 ks observation, the spectrum will be accurately measured up to 40 keV. Similar data will be obtained along the jet itself.

## 2.2 Galactic Centre region

With a supermassive Black Hole, the closest to us, at its centre the Galactic nucleus is one of the most interesting regions for high energy astrophysics[7], allowing a link with more distant galactic nuclei which cannot be studied with as much details. Despite recent advances, in particular due to Chandra[8, 9], the puzzling quietness of this 2.6 million solar masses Black Hole is a challenge to theory, and the nature, origin, and even existence, of a hot diffuse component is a matter of strong debates. SIMBOL–X will cast new light on this rich and astrophysically important region. Its sensitivity and angular resolution will allow to monitor the high energy activity of Sgr A* from quietness to strong flares, as illustrated in Figure 3, and thus distinguish between the different models that are proposed for its variability, particularly between those invoking accretion instabilities[10] and those invoking jets[11]. Similarly, the high energy diffuse emission will be mapped, and simulations have shown that its spectrum will be accurately measured up to more than 50 keV on scales smaller than 1 arcmin. This will allow an unambiguous characterization of this non-thermal emission, and give unique clues for unraveling its origin, for which a number of suggestions are competing (as e.g. Low Energy Cosmic Rays[12] or quasi-thermal electrons close to acceleration sites[13]). Finally, the SIMBOL–X hard X–ray map should also bring unique informations regarding the nature of the unidentified EGRET source 2EG J1746-2852, and in particular tell us if it is associated with SgrA East as proposed by Melia et al.[14].

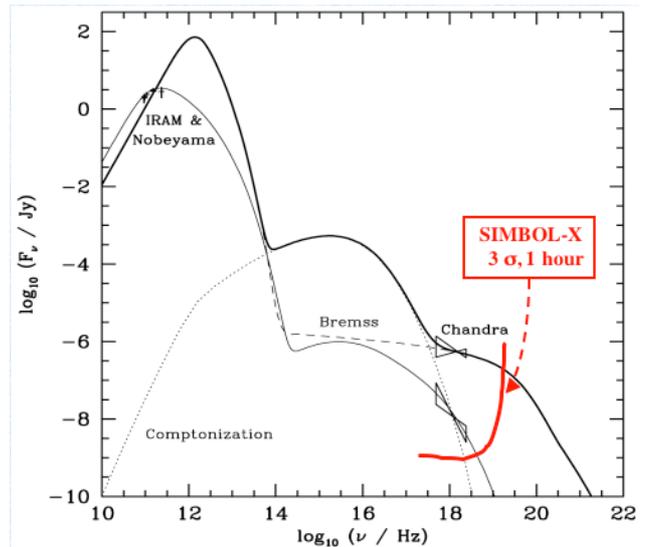

Figure 3 : Sensitivity of SIMBOL–X compared to the quiet and large flare X–ray flux level of SgrA*, with model predictions from Liu and Melia[10].

## 2.3 Obscured AGNs and the diffuse X–ray background

It is now largely admitted that the diffuse X–ray background (XRB) can be explained by a combination of absorbed and non-absorbed AGNs, as proposed initially by Setti and Woltjer[15], an hypothesis which has been remarkably confirmed by the recent deep field observations of XMM–Newton and Chandra[16, 17]. These observations have indeed resolved in sources most of the XRB flux below 10 keV, and found an important fraction of obscured AGNs and quasars. However, the peak of the redshift distribution of the detected obscured AGNs was surprisingly found at $z \sim 0.7$ whereas standard synthesis models predict a peak at twice that value. Solutions to this puzzle have been proposed in terms of a very quick evolution of obscured AGNs up to $z < 1$, and would link this population to that of starburst galaxies[18], thus with a strong implication on the infrared background and the understanding of the formation and evolution of galaxies.

It is important to note however that the XRB peaks around 30 keV, with more than 50 % of its energy flux in the 20–70 keV band, i.e. fully outside the band studied by XMM–Newton and Chandra on which these models are built. It is thus clearly essential to carry out a deep survey above 10 keV, to indeed find out what is the fraction of Compton thick sources contributing to the XRB, to have an unbiased sample for the cosmological evolution of AGNs and to estimate their contribution to the infrared emission of the Universe. We estimate that there are about one hundred AGNs per square degree in the 20–50 keV band above the limiting sensitivity of SIMBOL–X in this band ($\sim 6 \times 10^{-14}$ erg/cm$^2$/s in 100 ks), with half of them entirely new compared to XMM–Newton and Chandra. With arcsec coordinates for each source, SIMBOL–X will indeed resolve the XRB puzzle at its peak energy, and allow us to search for counterparts of the sources in other wavelengths.

### 2.4 Shock acceleration

SIMBOL–X will be ideally suited to map the non-thermal emission of the relativistic particles accelerated in Supernova remnants. This is in particular essential to find out the maximum energy that can be reached in these acceleration sites. In thermal remnants, such as Cas A, the synchrotron emission of the TeV electrons is mixed up with the thermal bremstrahlung, as well as the non-thermal bremstrahlung of less than 1 MeV electrons. This completely hampers a clean identification and measurement of the non-thermal component. At energies above 10 keV, the synchrotron component will dominate and will thus be unambiguously identified and mapped. In non-thermal remnants SIMBOL–X will allow to measure cut-off energies well above 10 keV if they are present. As an example, a simulation of the North-East emission of SN 1006 has shown that a $10 \times 10$ arcmin$^2$ region can be mapped in a total of 100 ks observing time (Figure 4), with acccurate spectra up to 30 keV measured in zones of 1 arcmin$^2$ of area. By comparing these maps and spectra to the radio maps, it will be possible to find out if the accelerated electrons have energies above several hundreds of TeV, something impossible to test today.

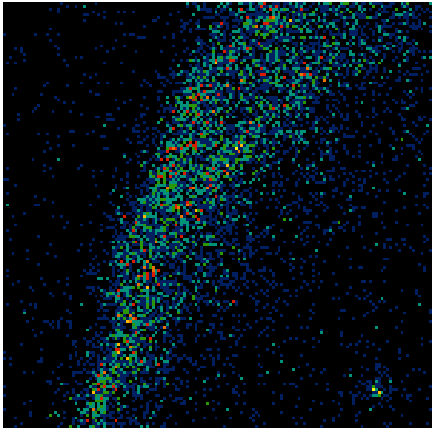

Figure 4 : Simulated SIMBOL–X image of SN1006 above 10 keV, obtained with a mosaic of pointings of 100 ks total observing time. The image, derived from actual XMM data, is $10 \times 10$ arcmin$^2$ wide.

## 3. MISSION CONCEPT

For its baseline design, SIMBOL–X is built using a classical Wolter I optics focusing X–rays onto a focal plane detector system. The gain in maximum energy is achieved by having a long focal length, of 30 metres, i.e. 4 times that of XMM–Newton mirrors. Since this cannot fit in a single spacecraft, the mirror and detectors will be flown on two separate spacecraft in a formation flying configuration, as sketched in Figure 5. The main scientific characteristics are given in Table 1, and correspond to the combination of mirror and detector systems described in the next sections. Here below, we give some details on a potential orbit, the formation flying requirements, as well as on the achieved sensitivity.

### 3.1 Mission requirements and orbit

The light weight mirror envisaged (~ 200 kg) and an anticipated modest weight of the focal plane (< 30 kg) ensure that this mission can be mounted on two spacecraft of the "mini-satellite„ class, and can be launched together into the operational orbit. The necessity to have a stable image quality, as well as to keep the full field of view inside the detector area, dictate the requirements on the formation flying stability. For the 30 arcsec HEW, 30 m focal length, nominal mirror, and a detector diameter slightly larger than the FOV these requirements are a position stability of ± 1 cm both along the telescope axis and perpendicular to it. As this is larger than the mirror PSF width (4.4 mm), in addition to these stability requirements there is a requirement on the knowledge of the position of the telescope axis whose movements must be constantly measured within 0.5 mm for a good oversampling of the 30 arcsec PSF.

The altitude of the SIMBOL–X orbit is chosen to be high enough to ensure that : i) the radiation belts induced background is as low as possible, and ii) the best compromise is found between the pertubations (gravity gradient) and the consumables needed to control a spacecraft on a forced orbit. The second requirement places the limit well above well shielded low Earth orbits, while experience from XMM–Newton shows that at the XMM–Newton inclination the radiation belt background disappears fully only above ~ 75,000 km.

These requirements form the basis of a mission study which is currently being conducted at the CNES French space agency, for CNES internal purposes. Besides allowing long uninterrupted observations, a circular orbit was preferred to simplify the formation flying control. The chosen altitude is 91,000 km. This study, which will be published elsewhere, demonstrates that the two spacecraft, fully equipped with propulsion systems (FEEPs) and ranging systems (RF and optical metrology) allowing to meet the formation flying requirements and enough consumables for a full 2 years mission with typical 100 ks long observations, would have a total mass of around 1.5 tons. The mirror spacecraft (master) will be on the circular orbit, while the detector spacecraft (slave) orbit will be forced. This study has also demonstrated that the spacecraft could be launched together and put directly in the right orbit by a Soyuz rocket equipped with a Fregat upper stage.

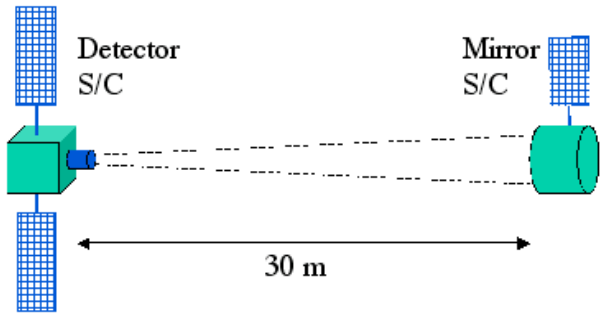

Figure 5 : Sketch of the SIMBOL–X configuration in flight

| | |
|---|---|
| Energy range : | 0.5 – 70 keV |
| Energy resolution | 130 eV @ 6 keV  [SDD] |
| | 1 %   @ 60 keV  [CZT] |
| Angular resolution | < 30 arcsec |
| Localisation | < 3 arcsec |
| Field of View | 6 arcmin (50 % vignetting) |
| Effective area | > 550 cm$^2$   E < 35 keV |
| | 150 cm$^2$  @ 50 keV |
| | 10 cm$^2$   @ 67 keV |
| Sensitivity | 5 10$^{-8}$ ph/cm$^2$/s/keV  < 40 keV |
| | (5 σ, 100 ks, ΔE = E/2) |

Table 1 : Main SIMBOL–X characteristics

### 3.2 Sensitivity

In order to estimate the sensitivity of SIMBOL–X, we have used the background spectra studied in the context of Constellation–X. This mission is designed for an L2 orbit, with particle fluxes similar to the high circular orbit of SIMBOL–X, and has an actively shielded Cd(Zn)Te detector, as SIMBOL–X. Ramsey[19] has published a detailed analysis of the detector background spectrum, consisting of a diffuse sky component and of an internal instrumental background. The SIMBOL–X background spectrum was calculated by adding the Constellation–X internal background to the diffuse sky component[20] correctly scaled to the relevant SIMBOL–X parameters (effective area, and required extraction region).

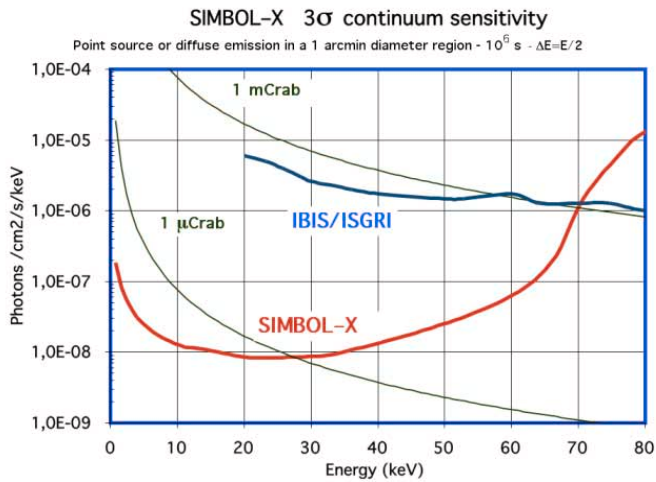

Figure 6 : SIMBOL–X broad band (ΔE = E/2) sensitivity to point sources, for a 1 Ms observing time and a 3 σ detection. The IBIS/ISGRI sensitivity is from Lebrun *et al.*[35]. The milli- and micro-Crab levels have also been indicated.

With this background spectrum, we have then derived the sensitivity curve shown in Figure 6 for a point source detection at the 3 σ level, in 1 Ms of observing time for comparison with INTEGRAL sensitivity. An extraction region of 1 arcmin diameter was assumed for the source photons, i.e. two times the HEW of the PSF. This sensitivity curve also describes the sensitivity of SIMBOL–X to diffuse emission on the scale of 1 arcmin diameter. As expected for an X–ray focusing telescope, the SIMBOL–X sensitivity curve has an XMM–Newton or Chandra like shape but is displaced by about a decade in energy. SIMBOL–X is more than 100 times better than INTEGRAL instruments below ~ 50 keV, and has a sensitivity equivalent to IBIS/ISGRI at ~ 70 keV.

### 4. MIRROR DESIGN

Soft X-ray telescopes like XMM–Newton and Chandra are based on grazing incidence Wolter I nested mirror shells. They exploit the total reflection phenomenon, which is characterized by a very large reflectivity at grazing angles until a cut-off angle α$_c$ (the so called "critical angle"), beyond which the reflectivity falls rapidly down to almost zero. In turn,

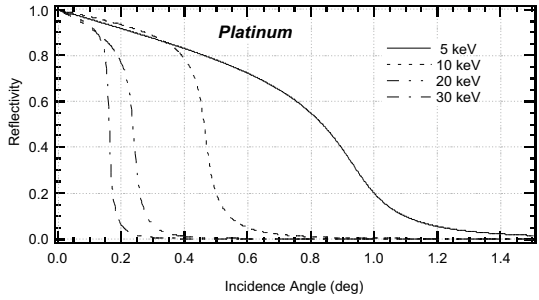

Figure 7 : Theoretical reflectivity profiles versus the incidence angles at 4 different energies for Platinum.

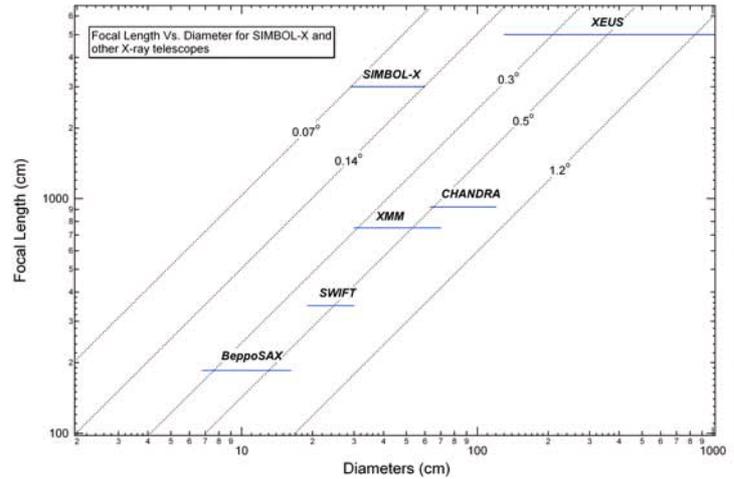

Figure 8 (right) : Focal Lengths versus Diameters for SIMBOL–X and other X–ray missions. We also have reported, for comparison, the corresponding incidence angles for reflection. As can be seen, SIMBOL–X is characterized by very shallow angles but, thanks to the very large focal length, the external aperture is just slightly smaller than XMM.

the critical angle linearly depends on the square root of the reflecting material density and on the inverse of the photon energy. This effect can be well appreciated in Figure 7 where, as an example, the theoretical reflectivity profile versus the incidence angle of a large density material as Pt is shown at different incidence angles and for different photon energies. Beyond 10 keV the useful angles for reflection become very small, determining, as a consequence, a strong limitation of the available collecting area for optics with usual focal lengths ($\leq$ 10 m). This is the main reason for explaining the difficulty in the realization of hard X-ray telescopes.

**4.1 Nominal mirror design**

A possible way-out to this problem is the use of multilayer-coated mirrors that, thanks to the Bragg diffraction phenomenon, are characterized by a larger reflectivity range than single layers (the gain is about a factor 3 with respect to Au[21]). It is already foreseen to make use of this solution in the realization of future Hard X-ray telescopes like, e.g., Constellation-X/HXT[22]. However it presents the drawback of a relatively complex and time-consuming manufacturing time, related to the multilayer films depositions. In addition, as firstly observed by Weisskopf *et al.*[23], the multilayer advantage is diminished by a relatively low efficiency after two reflection bounces (as it happens in the case of Wolter I mirrors). The use of conventional single-layer mirrors with very shallow reflection angles is a much easier way to be pursued from the point of view of the optical realization, provided that the problem of small apertures can be solved by the use of focal distances much larger than usual. This has been the criteria adopted to drive the design of the SIMBOL–X telescope (see Figure 8). Thanks to the long focal length (30 m), the aperture diameter is very large, similar to that of an XMM mirror module. On the contrary, the reflection angles are much smaller than XMM, allowing us to get a high effective area up to 70 keV.

| | |
|---|---|
| *Max/Min Diameter* | 600/290 mm |
| *Focal Length* | 30000 mm |
| *Mirror Heigth* | 600 mm |
| *Configuration* | Wolter I |
| *Number of mirror modules* | 1 |
| *Number of Mirror shells* | 100 |
| *Reflecting coating* | Pt |
| *Min/Max inc. angles* | 0.07°/0.142° |
| *Material of mirror walls* | Ni |
| *Min/Max wall thickness* | 0.12/0.30 mm |
| *Mass of the mirror module* | 213 Kg |
| *Field-of-View (50 % vignetting)* | 6 arcmin |
| *Expected resolution (HEW)* | 30 arcsec |
| *Effective area @30 keV* | 580 cm$^2$ |

Table 2. Main parameters of the SIMBOL–X mirror.

It should be noted that the SIMBOL–X field of view (~ 6 arcmin in diameter at 50 % vignetting) is however smaller by about a factor 2 than what could be achieved with multilayers. The main parameters of the optical module are reported in Table 2. Based on the experience acquired with the Beppo–SAX, Jet–X/SWIFT, and XMM–Newton mirrors, the mirrors will be made following the Nickel electroforming replication method[24]. This offers the advantage to provide a high throughput but preserving high imaging performances.

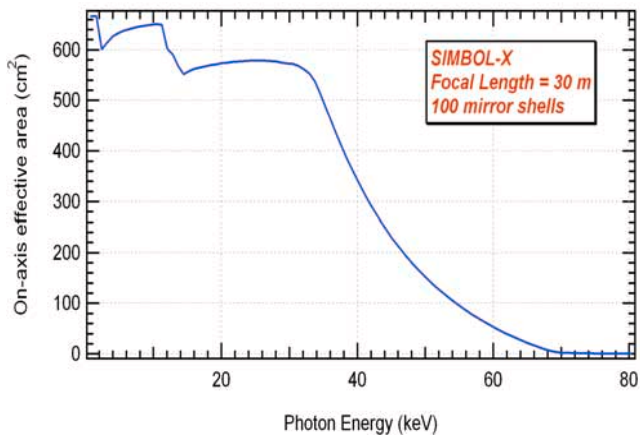
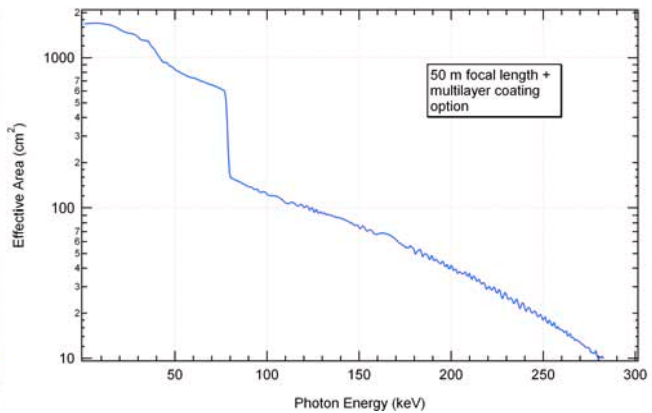

Figure 9 : Effective area for nominal SIMBOL–X design.

Figure 10 : Theoretical effective area of SIMBOL–X assuming a 50 m focal length and a Pt/C multilayer coating.

The reflecting material will be Platinum. This coating presents a slightly larger (~ 6.5 %) reflection window than Gold and, as already proven by experimental tests, it can be evenly used for successfully pursuing the Ni replication process. In order to get a mass of the optical system as small as possible (for being compatible with small mission class), in SIMBOL–X the thickness-to-diameter ratio for mirror walls is diminished by a factor 3.4 with respect to XMM. As a consequence the angular resolution will tend to be worse, but it is expected to be anyway better than 30 arcsec Half-Energy-Width, in agreement with experimental proofs already obtained[25]. Figure 9 shows the effective area as a function of energy. It is over 600 cm$^2$ at low energy and has roughly a constant value up to about 35 keV, before starting to decrease and fall below one cm$^2$ above 70 keV.

### 4.2 Alternative improved designs

The above baseline SIMBOL–X design is conceived in the framework of a "fast and simple„ project, in particular for the aspect concerning the mirror production. This is based on the Ni replication method for single-layer reflecting coating that can be considered a consolidated technology "ready off the shelf„. However the adopted formation flying approach gives us a very large flexibility in terms of possible improvements of the telescope performances, that we will be considering. The most straightforward are improvements in terms of effective area and angular resolution, in addition to the extension of the operational range to higher energy and the improvement of the field of view by the use of multilayer coatings. Of course in this case we would have to pay in terms of weight, complexity and, probably, the launch date would have to be delayed with respect the present timeline. However the scientific return could be greatly increased. As an example, we have here considered the following design : a 50 m focal length telescope made of 140 shells with diameters ranging from 100 cm (of the largest mirror) down to 30 cm (of the smaller one), with incidence angles between 0.142$^o$ and 0.044$^o$. We assumed segmented mirror shells made of thin glass 80 cm height, with a fixed wall thickness of 1 mm. The mirror weight is 640 Kg, including the mechanical structure. The manufacturing technique to be used in this case should be that being developed by the OAB and MPE for the XEUS mirrors, which is based on a thermal shaping of the glass panels plus a successive high-precision robotic polishing[26]. This approach is very promising and one could expect to achieve imaging capabilities as high as 10 arcsec HEW or better with very thin glass plates after a few years of the development activities now in course. As a reflecting coating we assumed for a simulation Pt/C multilayer made of 250 bilayers, whose parameters were defined after just a few iteration optimization. The theoretical effective area is shown in Figure 10. The high improvement of the effective area and the appreciable extension of the operative range are evident.

## 5. FOCAL PLANE DESIGN

The focal plane detector system will combine a Silicon low energy detector, efficient up to ~ 20 keV, on top of Cd(Zn)Te high energy detector. They will be surrounded by an active CsI anticoincidence shield. The detector diameter will be of 6 cm, slightly more than nominally required by the mirror Field of View. Both detectors are spectro-imagers, and will have a pixel size of 500 μm maximum, which will allow a sufficient oversampling of the mirror point spread function (4.4 mm diameter HEW). Both detectors can also be read out at high speed, a fact which will be used to efficiently reduce the high energy background by using them in the anticoincidence scheme in addition to the CsI guard counter. The two imager detectors are described below.

**5.1 The DEPFET Active Pixel Sensor low energy spectro-imager**

**5.1.1 DEPFET integrated amplifier**
The integrated detector-amplifier structure DEPFET (DEpleted P channel Field Effect Transistor) consists of a p-channel field effect transistor on a high resistivity n-type silicon bulk[27, 28]. The transistor may be either a JFET, or a MOSFET of enhancement or depletion type. The bulk is fully depleted by a reverse biased backside diode thus creating a potential minimum for electrons close to the surface. An additional deep n-doped region enhances the depth of the potential minimum and confines it in the lateral direction to the extent of the FET gate (Figure 11).

Each electron released in the depleted volume by thermal generation or by the absorption of ionizing radiation has to drift to the potential minimum and enhance the transistor current by inducing an additional positive image charge inside the FET channel. Thus the DEPFET's current is a function of the amount of charges in the potential minimum, and its measurement yields information about the energy absorbed in the depleted volume. To express the current steering function of the stored electrons the potential minimum is called "internal gate„. Its measured sensitivity is 200 pA per electron[29], thus the signals generated by X-ray photons with energies above 100 eV are an easily measurable quantity.

The internal gate exists, i.e. electrons are collected and stored in it, regardless of a current flowing in the channel of the DEPFET or not. The transistor current may be turned off during integration and switched on by the external gate only for the signal readout, thus minimizing power consumption. Its geometric size and doping concentration are dimensioned to store up to a total of 100.000 signal electrons. Unlike a conventional detector-preamplifier system, the DEPFET is free of interconnection stray capacitance and the overall capacitance is minimized. Depending on leakage current and signal rate the device must be reset periodically by emptying the internal gate. This is done by applying a positive voltage pulse to an adjacent n+ doped "clear„ contact acting as drain for electrons. The clear contact is separated from the detector bulk by a deep p-implanted well (Figure 11).

The noise characteristics of single DEPFET devices of the MOSFET type have been evaluated by spectroscopic measurements. The DEPFET was configured in a source-follower circuit, and the data have been taken by a commercial spectroscopy system. Spectra of a radioactive $^{55}$Fe source have been recorded and energy-calibrated with the use of the

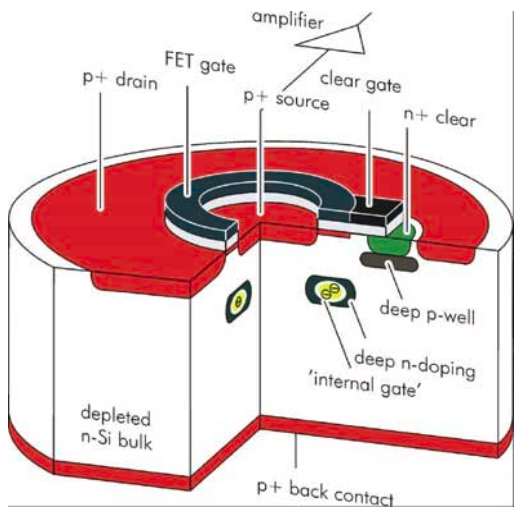
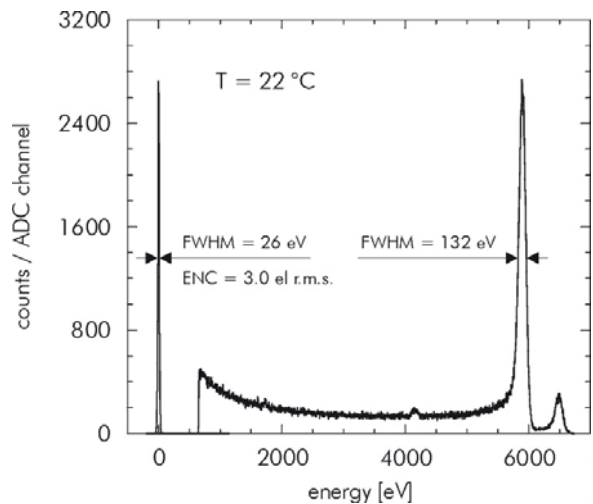

Figure 11 : DEPFET in circular geometry. Electrons generated by the absorption of ionizing radiation will drift to the potential minimum of the "internal gate„ and enhance the transistor current by inducing an additional positive image charge inside the FET channel. Applying a positive voltage pulse to the clear contact and to the MOS clear gate resets the device.

Figure 12 : Spectrum of a radioactive $^{55}$Fe source recorded with a single DEPFET at room temperature. The electronic noise contribution expressed in equivalent noise charges (ENC) is only 3 el. r.m.s. at room temperature.

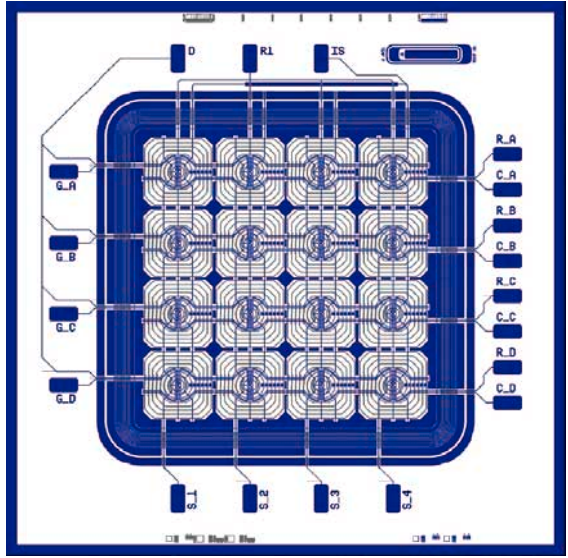

Figure 13 : Layout plot of a 4 × 4 Active Pixel Sensor prototype with 1 mm² pixel size. Each pixel consists of a square-shaped Silicon Drift Detector with an integrated DEPFET in the centre. The sensor area is surrounded by guard rings and bond pads.

Mn Kα line at 5.9 keV and the noise peak. The noise peak has been measured separately while the detector was not exposed to radiation. From the width of the noise peak an equivalent noise charge of 3 el. rms at room temperature has been deduced (Figure 12). In a single-pixel measurement incomplete charge collection caused by charge splitting at the pixel borders cannot be discriminated or corrected for. Therefore the spectrum of Figure 12 shows a pronounced low energy background, and the width of the Mn K lines is broadened by this effect as well.

### 5.1.2 DEPFET-based Active Pixel Sensor

The matrix-like formation of DEPFETs with common back contact results in a backside illuminated Active Pixel Sensor (APS) with a unity fill factor. The thickness of the depleted bulk is 500 μm, increasing the quantum efficiency at high energies to 96 % at 10 keV and 45 % at 20 keV. On the low energy side the thin entrance window technology results in quantum efficiency values above 90 % down to 300 eV[30].

All DEPFETs in the pixel sensor have a common drain contact, while the gates, clear contacts, and clear gates are connected row-wise, and the source contacts, i.e. the signal lines, are connected within each column. To selectively activate a row of pixels for readout or reset, control chips of the SWITCHER type[31] in a high-voltage CMOS technology are connected to the gate and clear lines. The readout is done in parallel for all pixels of one row by a preamplifier-filter-multiplexer chip of the CAMEX type[32].

As the individual pixels are randomly accessible, the APS has a high degree of flexibility in the choice of readout modes, depending on the object and the scientific goal of an observation: In full frame mode the whole sensor area is read row by row. With the exception of one active row, all pixels are turned off and are in integration mode, thus keeping dead time short and power consumption low. The processing time for one row is of the order of a few micro-seconds. In window mode only selected regions of interest (ROIs) that may have arbitrary rectangular shapes and sizes are read out, while the other pixels are suppressed. Time variations of fast transients can be observed in timing mode, i.e. a selected ROI is read at maximum speed but with reduced energy resolution. Any combination of the above specified readout modes or mixed mode applied to dedicated regions of the sensor area is possible.

The minimum pixel size of a DEPFET-based APS is $50 \times 50$ μm², given by the current technology rules. To increase the pixel area without running into the problem of weak lateral fields and long carrier collection times, the DEPFET is placed in the centre of a Silicon Drift Detector (SDD)[33]. A SDD consists of a depleted volume in which an electric field with a strong component parallel to the surface drives signal charges towards a small sized collecting electrode. The SDD principle combines a large sensitive area with a small output capacitance. That way any pixel size from $50 \times 50$ μm² up to $1 \times 1$ cm² can be adjusted to the resolution of the telescope without losing the DEPFET's energy resolution. Figure 13 shows a layout plot of a $4 \times 4$ APS prototype based on the DEPFET–SDD combination with $1 \times 1$ mm² pixels which is currently in production.

### 5.2 The pixellated Cd(Zn)Te high energy spectro-imager

The high-energy camera of SIMBOL–X is based on a modular design. The modules will constitute a mosaic of pixellated arrays of Cd(Zn)Te (CZT), room temperature semiconductor detectors, working in the range of 5 keV up to 70 keV. The dynamic range of the low energy and the high energy detectors will overlap. The design of the high energy detector takes into account the scientific needs in terms of high spatial and high spectral resolution. This leads to invent a hard X-ray camera with a high density of pixels and capable of excellent spectral performances up to 70 keV. On the other hand, the design integrates the space conditions specificities (background rejection, radiation damage, design robustness, low power and low mass, passive cooling) in addition with the new formation flying conditions (relative positions of the two spacecraft and camera size).

### 5.2.1 The CZT basic component of the high energy layer

The basic component of the high energy layer of the SIMBOL–X focal plane is based on the use of Cd(Zn)Te and has to work between 5 keV and 70 keV. Thanks to relatively high atomic numbers ($Z_{Cd} = 48$, $Z_{Te} = 52$), which gives a dominant photoelectric probability up to 250 keV against Compton scattering interaction process, and thanks to a high-density (~ 6 g/cm$^3$), which gives a good stopping power, CdTe shows high quantum efficiency suitable for detection of photons in the typical range of 5 to 500 keV[34]. Moreover, one can consider 100 % total efficiency up to 80 keV for a 2 mm thick detector (Figure 14), which is well suited to the SIMBOL–X application. The Compton scattering probability at 80 keV is roughly 10 % and the escape probability of the scattered photon in a 1 cm$^3$ crystal is almost negligible.

Furthermore, its wide band-gap leads to a very high resistivity ($10^9$ to $10^{11}$ Ω cm) and allows operations at room temperature. Finally, the intrinsic spectral resolution, simply by taking into account statistical fluctuations of electron-hole pair creation, compares well with that of Germanium (400 eV and 620 eV FWHM at 100 keV respectively for Ge and CZT). This makes CdTe a very promising material since manufacturing difficulties and signal processing problems are overcome. Current manufacturing technologies of CdTe and Cd(Zn)Te allow working with crystal volumes up to 1 or 2 cm$^3$. On the other hand, CdTe has demonstrated its capabilities in space through the success of the ISGRI detector on board the *INTEGRAL* satellite[35].

Pixel arrays of rather large dimensions (1 to 4 cm$^2$), equipped with small pixel size patterns are capable of impressive spectral performances. The segmented anode pattern is well adapted to charge collection in CdTe / Cd(Zn)Te. Actually, in segmented detectors, contrary to standard coplanar detectors, the induced current no longer only depends on the distance that carriers travel (penetration depth), but also on their proximity to the sensing electrode (anode)[36, 37, 38]. This is the "near field effect„ (or "small pixel effect„) and it can be exploited to define segmented anode geometry with small pixel in order to distribute the weighting potential close to the anode. This helps to remove the hole contribution to the signal. When a charge packet is generated in such a device, it induces a surface charge on any nearby electrodes. As the packet drifts along the applied electric field lines, the influence of carriers motion is equally distributed on every pixel until the packet moves very close to the one in front of it, where the weighting potential is strong. When charge carriers move within a hemispherical region around the pixel, approximately of the size of the lateral pixel size, a strong signal formation occurs in the pixel electrode. This small pixel geometry is good for signal formation because the signal is induced lately and no ballistic deficit takes place. Consequently, the recorded lines are almost symmetric and present a nice quasi-Gaussian shape, easy to handle. On the other hand, the benefit of very fine pixel size is that the total crystal dark current and capacitance is shared over many channels and very low noise can be achieved.

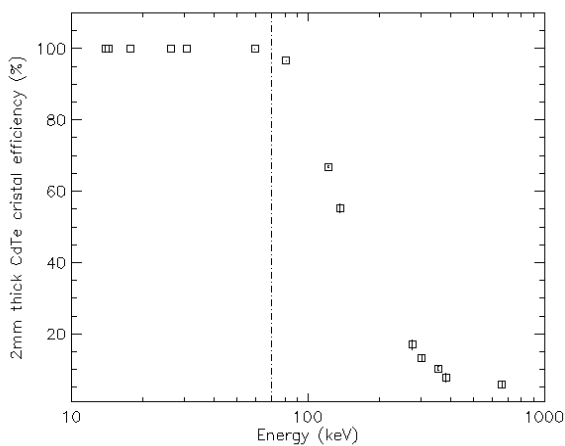
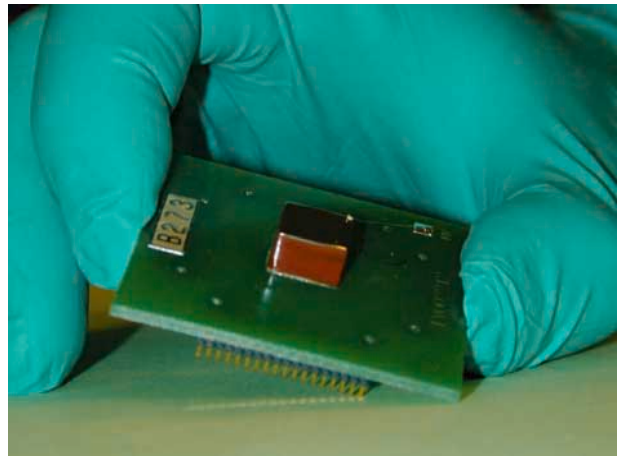

Figure 14 : Absorption efficiency of a 2 mm thick CdTe crystal.

Figure 15 : Cd(Zn)Te pixellated crystal under test in Saclay.

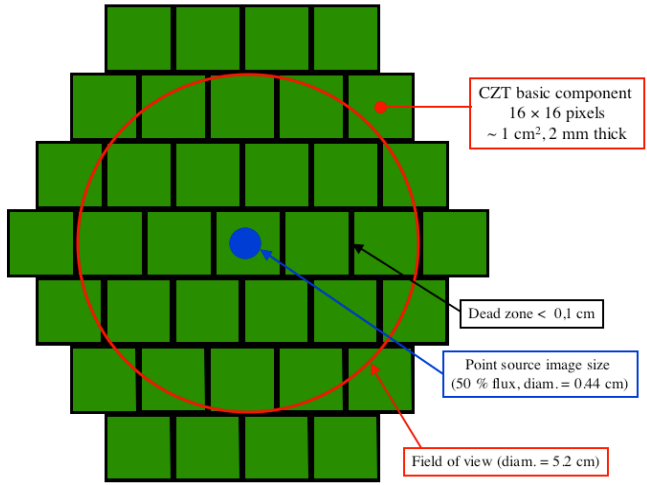

Figure 16 : Basic shape of the SIMBOL–X high energy focal plane made of a mosaic of 37 elementary CZT arrays. The design will be made in order to limit the dead zones.

The CZT layer of the SIMBOL–X mission is based on the use of such devices. The pixel size, due to the PSF of the instrument, has to be about $500 \times 500$ μm$^2$. Using 2 mm thick detectors, the devices promise for very high spectral performances at a moderately negative temperature ($\sim -20°C$). As an example, the HEFT design leads to 550 eV FWHM at 60 keV with 2 mm thick CZT and $600 \times 600$ μm$^2$ pixels at $-25°C$[39]. This is possible thanks to a very low noise front-end integrated electronics, directly bumped onto the anodes. Our aim for the SIMBOL–X mission is to provide a pixel array with such spectral performances at least. To succeed, we are developing a new analogue front-end electronics which will be directly hybridized on our crystals. We plan to obtain an equivalent noise charge of less than 60 electrons rms. The chip will handle 256 pixels in a single circuitry. A digital part of the ASIC will be in charge of the addressing and gain, low threshold and high threshold tuning. The ASIC is designed in a sub-micron technique AMS 0.35 μm. The design will be tolerant to radiation dose compatible with the orbit of SIMBOL–X and its life-time. The area of the chip will suit exactly the surface of the detector in order to avoid dead spaces between arrays in the mosaic assembly of the focal plane.

The basic shape of the detector will be a little less than 1 cm$^2$ crystal, 2 mm thick equipped with $16 \times 16$ pixels of $500 \times 500$ μm$^2$ separated by a 50 μm gap. All of the 256 pixels will be surrounded by a 100 μm guard ring. The guard ring itself is about 100 μm from the edge of the part. Thicker geometries up to 10 mm will be tested also for spectral response purpose. The detectors under study in CEA Saclay at the moment are provided by eV-products Company (Figure 15).

**5.2.2 The modular focal plane**
In order to guarantee a high level of reliability and easy mounting, the design of the high energy detector plane will be fully modular. Each basic element of CZT will be mounted in thin mechanical cells, limiting dead spaces. The signals will be extracted below the assembly and routed to integrated ADC's, close to the detector layer. The control electronics of the overall detector will be able to handle multiple events and will provide accurate tag time to perform coincidences with the low energy detector and the veto sub-systems surrounding the camera.

Basically, the high energy detector will be a mosaic of 37 devices which leads to 9472 independent channels. The shape of the focal plane is approximately a disk and the central element is presented to the on-axis focal point to avoid dead zones in the image centre (Figure 16). This detector plane will include the full field of view of the telescope even when allowing for a ~ 1 cm jitter of the two spacecraft relative positions.

## 6. CONCLUSIONS

The SIMBOL–X mission will provide an unprecedented sensitivity and angular resolution in the high energy domain, enabling to solve a number of standing questions in the non-thermal universe. In its baseline design, SIMBOL–X uses existing technologies both for the optics and the detectors. It relies on the yet non tested formation flying concept, but with requirements that can be met within existing technologies. SIMBOL–X will be first to use in space this "open tube„ telescope in the X–ray domain, a configuration that has been selected for XEUS for obtaining tens of square metres of effective area. As described, SIMBOL–X might also make use of the XEUS mirror technology development.

The SIMBOL–X proposal was presented to the CNES French space agency in 2002, as a multilateral mission, in the context of its former quadri-annual selection process. The proposed launch date is around 2010. The proposal was recommended for a start of phase A study by the CNES Astrophysics advisory committee.

## ACKNOWLEDGMENTS

We are profoundly indebted to all the people who have actively participated to the elaboration of the SIMBOL–X proposal, especially within the Service d'Astrophysique of the CEA, and who are too numerous to be cited here. We thank Vincenzo Cotroneo from OAB for performing the multilayer calculations in the 50 m focal length option. Philippe Ferrando also warmly acknowledges the CNES PASO team, led by Karine Mercier, for their SIMBOL–X mission study, and for sharing part of their results before publication.